\documentclass[a4paper,11pt,fleqn]{article}

\usepackage{mathptmx}
\usepackage{graphicx}
\usepackage[english]{babel}
\usepackage[utf8]{inputenc}
\usepackage[numbers,sort&compress]{natbib}
\usepackage[colorlinks,citecolor=blue,linkcolor=blue, urlcolor=blue]{hyperref}
\usepackage{xcolor}
\usepackage{textcomp}
\usepackage{amsmath,amsfonts}
\usepackage{amssymb}
\usepackage{graphicx}
\usepackage{float}
\usepackage{lineno}
\usepackage{times}
\usepackage{epsfig}
\usepackage{epstopdf}
\usepackage{color}
\usepackage{lscape}
\usepackage{rotating}
\usepackage{multirow}
\usepackage{siunitx}
\usepackage[font={small,sf,stretch=0.9},labelfont=bf]{caption}

\begin{document}

\begin{center}

 {\Large{\textbf{Anomalous thermal and elastic properties of an epitaxial NiTi film \\[2mm] exhibiting R-phase}}}

 \bigskip
 {\large
 	K.~Rep\v{c}ek$^a$,
	T.~Grabec$^b$*,
	D.~Mare\v{s}$^b$,
	P.~Stoklasov\'{a}$^b$,
	P.~Sedl\'{a}k$^b$,
	J.~Ku\v{s}n\'{i}r$^b$,
	P. Ve\v{r}t\'{a}t$^c$,
	O. Heczko$^c$,
	S.~F\"{a}hler$^d$,K.~L\"{u}nser$^{e,f}$,
	H.~Seiner$^b$
	}

 \bigskip \emph{
 $^a$ Faculty of Nuclear Sciences and Physical Engineering, Czech Technical University in Prague, Czech Republic \\
 $^b$ Institute of Thermomechanics, Czech Academy of Sciences, Czech Republic\\
 $^c$ FZU -- Institute of Physics,  Czech Academy of Sciences, Czech Republic\\
 $^d$ Helmholtz-Zentrum Dresden-Rossendorf, Dresden, Germany \\
 $^e$ Institute for Energy and Materials Processes, Universität Duisburg-Essen, Germany \\
 $^f$ Research Center Future Energy Materials and Systems (RC FEMS), Research Alliance Ruhr, Bochum, Germany}

 \bigskip \emph{*corresponding author; email: tomas.grabec@it.cas.cz}

\end{center}
\bigskip
\paragraph*{Abstract}
Shape memory alloys like NiTi are at the core of emerging thermal management applications, including elastocaloric refrigeration, thermoelastic harvesting, and latent heat storage.
Most of these applications benefit from a small scale due to the accelerated heat exchange, but obtaining precise functional properties of films is challenging.
Here we demonstrate that transient grating spectroscopy (TGS) enables characterization of elastic coefficients and thermal diffusivity of a \SI{3}{\micro\metre} thick epitaxial NiTi film during a thermally induced phase transformation.
The in-situ measurement of a complete austenite$\rightarrow$R-phase$\rightarrow$martensite$\rightarrow$austenite temperature cycle reveals that the elastic properties exhibit a crossover of the shear moduli (from $c^\prime < c_{44}$ in austenite to $c^\prime > c_{44}$ in martensite) and that the thermal diffusivity changes by a factor of \num{4.5} between the R-phase and austenite.
This dramatic change, together with the absence of hysteresis between the R-phase and austenite, makes NiTi a promising material candidate for thermal switches.
The results indicate that the change in thermal diffusivity originates from an anomalous heat capacity of the R-phase.
Furthermore, our TGS study provides temperature-dependent thermal and elastic properties required for simulating thermal management microsystems using this material.

\bigskip
\section{Introduction} 											\label{sec:Intro}
	Beyond medical and actuator applications, shape memory alloys drive many emerging applications in thermal management.
	These applications include elastocaloric refrigeration~\cite{Bonnot2008,Tusek2016}, thermoelastic energy harvesting~\cite{Goldstein1978,Neumann2025,Zou2026,Khosravi2025}, latent heat storage~\cite{Hite2025}, and morphable thermal protective clothing~\cite{Pan2023}.
	Most of these applications benefit from small size that accelerates heat exchange.
	This increases power density accordingly, motivating, for example, the development of elastocaloric microcoolers based on thin films~\cite{Bruederlin2018}.

	Another component required for thermal management is thermal switches, which can switch their thermal diffusivity between a low and high state, as reviewed by {Klinar et al.}~\cite{Klinar2021}.
	These thermal switches are of particular interest for cyclic thermal engines as they allow connecting or disconnecting the engine in alternation to the hot or cold reservoirs.
	Two properties are beneficial for thermal switches: They should switch quickly and exhibit a high ratio between both conducting states.
	Mostly fluid or mechanical actuation is used for thermal switches, including those utilizing smart active materials~\cite{Benafan2013,Simmers2007,Voigt2018,Chen2026SMST}, but they have the drawback of moving parts, limiting cycle count~\cite{Klinar2021}.
	Solid-state thermal switches do not have this drawback, as the functionality is built within the material.
	VO$_2$, which exhibits a metal-insulator transition, is one of the most promising materials.
	While the first report demonstrated changes in thermal conductivity of \SI{60}{\percent}~\cite{Oh2010}, now up to \SI{250}{\percent} is obtained~\cite{Hamaoui2019}.
	Shape memory alloys also exhibit this property, and for a transition between austenite and martensite in NiTi values between \SI{94}{\percent}~\cite{Garcia-Garcia2014} and \SI{100}{\percent} were obtained~\cite{Faulkner2000}.
	The high thermal conductivity of metallic shape memory alloys is beneficial for fast heat transfer.
	Thin films are then particularly suited for thermal microsystems where high cycling frequencies and compact footprints are required. For these, solid-state thermal switches will be the only choice.

	In all thermal management applications described above, precise knowledge of thermal properties in thin films is also crucial.
	A variety of methods have been developed for measuring thermophysical properties of films.
	Electrical techniques, most notably the $3\omega$ method, have proven robust for measuring the thermal conductivity of thin films and are now well-established~\cite{Jain2008,Jacquot2010}.
	However, they necessitate the microfabrication of metallic heater-thermometer structures on the sample, which can be complex and may influence the measurement~\cite{Yamane2002}.
	For this reason, non-contact optical techniques based on a pump-probe architecture, such as time-domain and frequency-domain thermoreflectance, have become dominant in recent decades~\cite{Zhu2010}, especially for measuring cross-plane heat flow~\cite{Jiang2017AdMat,Feser2012,Jiang2017}.

	Transient grating spectroscopy (TGS)~\cite{Maznev1998,Hofmann2019,Choudhry2021,Stoklasova2021} is an established contactless method for multiphysical characterization,
	capable of evaluating multiple material properties, such as thermal diffusivity~\cite{Wylie2025,Trachanas2025,Reza2020,Dennett2018,Kusnir2023}, elasticity~\cite{Repcek2024,Xu2022,Duncan2016}, radiation damage~\cite{Liu2025,Simmonds2025,Reza2022,Dennett2019,Dennett2018a,Short2015}, acoustic dynamics~\cite{Tobey2006}, phonon transport~\cite{Minnich2015,Vega-Flick2017}, hole transfer efficiency~\cite{Jeong2025} and exciton diffusive transport~\cite{Bi2025} in semiconductors.

	It is a laser-based ultrasonic method where two laser beams interfere to create a spatially periodic excitation pattern.
	For an opaque sample, this pattern acts at the surface as a thermoacoustic source, inducing dynamic surface ripples that combine transient thermal and acoustic gratings, as well as transient reflectivity changes.
	The method constrains the spatial characteristics in both propagation direction and wavelength, enabling detailed characterization of angular dispersion of properties.

	Micrometer-scale acoustic wavelengths make TGS an effective tool for probing the elasticity of thin-film systems, whether in freestanding or supported configurations~\cite{Rogers2000,Sermeus2014,Sermeus2015}.
	Notably, the technique was used for mapping thermal evolution of mechanical properties of a heterogeneous sputtered NiTi film, enabling identification of martensitic phase transition temperatures in various locations~\cite{Grabec2016}.
	The TGS method was also proven suitable for studying elasticity of epitaxial films made from strongly anisotropic material~\cite{Heczko2018}.

	Here, we employ TGS to investigate simultaneously the in-plane thermal diffusivity and elasticity of an epitaxial NiTi film across the martensitic phase transformation.
	Epitaxial NiTi thin films~\cite{Luenser2023,Luenser2024} offer the advantage of a well-defined crystallographic orientation.
	This is crucial for maximizing the anisotropic shape memory effect and for fundamental studies of the martensitic transformation without the influence of grain boundaries.
	Moreover, the films exhibit {faster cyclic response} compared to their bulk counterparts~\cite{Fu2004}.
	Epitaxial growth as a bottom-up approach also avoids several technological issues that affect NiTi in top-down manufacturing, including difficult machinability due to superelasticity or complicated melt-processing and composition control, both due to the high reactivity of titanium~\cite{Elahinia2012,Elahinia2016}.

	This study follows up on the work of \textit{Grabec et al.}~\cite{Grabec2024}, which used TGS to study the elastic properties of the epitaxial NiTi film in the two distinct states: high-temperature (austenitic) and low-temperature (martensitic).
	There, TGS detected multiple surface acoustic wave modes (SAWs), which enabled the full elastic tensor of both phases to be determined from a single-wavelength measurement.
	Here, we use this power of the TGS approach to monitor how the elastic properties change across the R-phase and martensitic transitions.
	In addition, we demonstrate the ability of TGS to capture the evolution of thermal diffusivity across these transitions.
	Together, these measurements provide new insight into the thermodynamics of the R-phase and martensite, and the suitability of NiTi as a solid-state thermal switch.

\section{Results} 								\label{sec:Results}

	\subsection{Film structure and phase transformation sequence}								\label{sec:FilmAndXRD}

	The studied NiTi thin film was epitaxially deposited on a single-crystalline MgO$(1\,0\,0)$ substrate via DC magnetron sputtering.
	Its thickness, determined by scanning electron microscopy on a focused-ion-beam cross-section, was \SI{3060}{\nano\metre}~\cite{Grabec2024}.
	To minimize lattice mismatch between the substrate and the NiTi film, a bilayer buffer system consisting of chromium (Cr) and vanadium (V) was employed (with a combined thickness of approximately \SI{75}{\nano\metre}).
	The epitaxial relationship between the austenitic B2 phase of NiTi, the buffer layers, and the MgO substrate is described by the crystallographic alignment MgO$(1\,0\,0)[0\,0\,1] \; || \; $V/Cr$(1\,0\,0)[0\,1\,1] \; || \; $NiTi B2$(1\,0\,0)[0\,1\,1]$
	 --	this means that the NiTi B2 unit cell is rotated by \SI{45}{\degree} about the surface normal relative to the MgO lattice.
	The sample had lateral dimensions of approximately \qtyproduct{10 x 10}{\milli\metre}, with the MgO substrate measuring around \SI{155}{\micro\metre} in thickness.
	For further details about the manufacturing process, see~\cite{Luenser2023}.

	\begin{figure}[tbhp!]
		\centering
		\includegraphics[width=\textwidth]{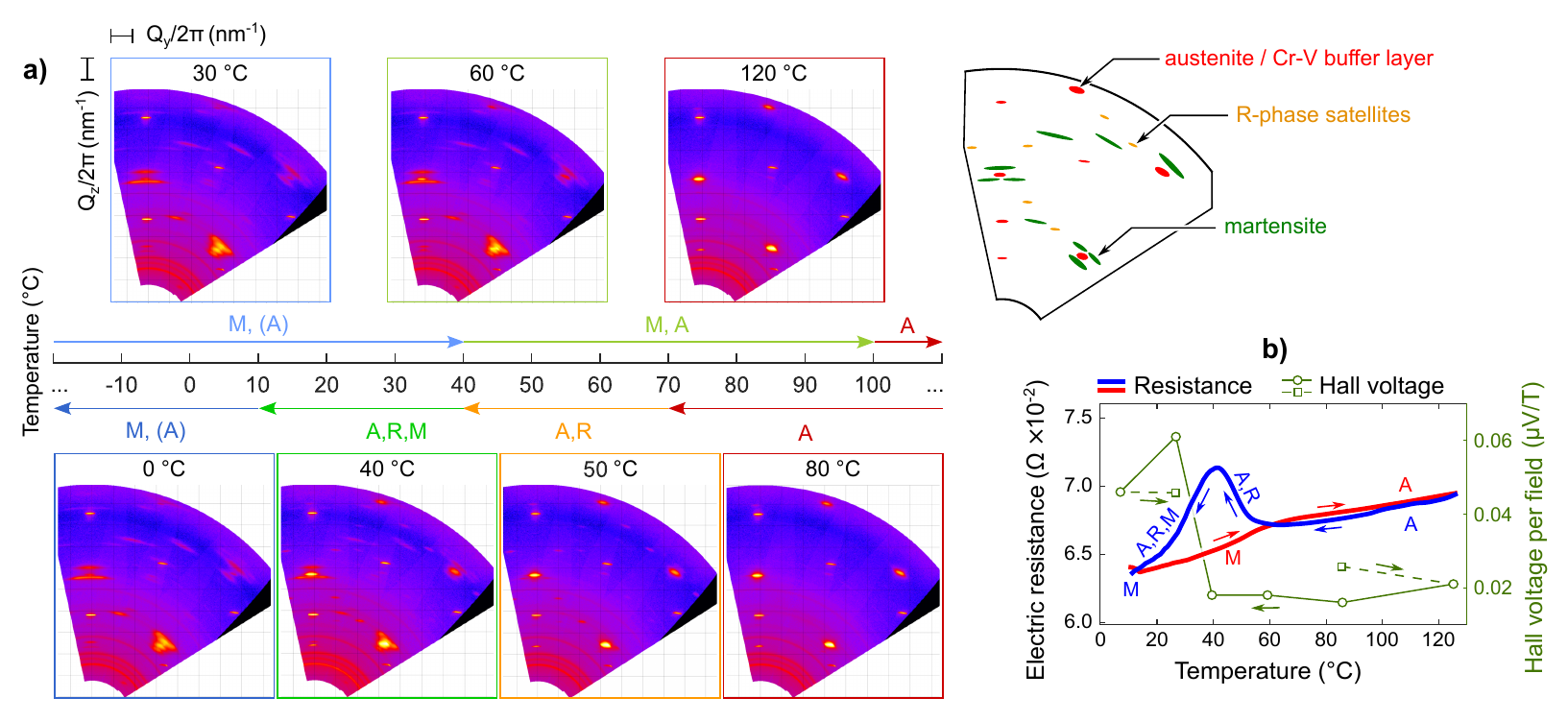}
		\caption{
			Phase transformation sequence of the studied NiTi film, characterized by a) XRD and b) electrical resistance and Hall voltage measurements.
			In a), X-ray maps at selected temperature are shown; maps above/below the temperature axis refer to the heating/cooling run, respectively.
			Phase reflections are identified in the right-hand panel; the $1/3\left<1 1 0\right>^*$ satellites characteristic of the R-phase are clearly visible during cooling but absent on heating.
			In b), the Hall voltage shows no anomaly at the R-phase transition but increases steeply at the onset of martensite formation.
			}
		\label{fig:XrayMaps}
	\end{figure}

	To determine the phase composition and transformation temperatures of the film, reciprocal space mapping X-ray diffraction (XRD) experiments were performed.
	The resulting 2D maps at selected temperatures are shown in Fig.~\ref{fig:XrayMaps}a).
	The heating and cooling runs were performed between \SI{5}{\degreeCelsius} and \SI{120}{\degreeCelsius} at a rate of \SI{1}{\degreeCelsius\per\minute}, with the XRD maps recorded every \SI{10}{\degreeCelsius}.
	The XRD measurements were conducted	using a Rigaku SmartLab diffractometer with Cu tube ($K\alpha_1$ = \SI{1.5406}{\angstrom}) in parallel-beam geometry and an Anton Paar DCS 500 domed cooling stage for in-situ temperature control.
	Analysis of the XRD maps showed that during the heating run from martensite, the austenite phase started to form at \SI{40}{\degreeCelsius}, with only faint traces of martensite remaining at \SI{80}{\degreeCelsius}.
	The transformation was completed at \SI{100}{\degreeCelsius}.
	During the cooling run, the R-phase started to form at \SI{70}{\degreeCelsius}, as indicated by the appearance of $1/3\left<1 1 0\right>^*$ satellite reflections~\cite{Otsuka2005}.
	These reflections strengthened upon further cooling to \SI{50}{\degreeCelsius}, where the martensite phase first appeared.
	At \SI{10}{\degreeCelsius}, the transformation to martensite was completed.

	The austenite $\rightarrow$ R-phase $\rightarrow$ martensite $\rightarrow$ austenite sequence deduced from the X-ray analysis was also confirmed by electrical resistance measurement (performed on an in-house built probe), as shown in Fig.~\ref{fig:XrayMaps}b).
	While the heating run showed a smooth transition from martensite to austenite, a typical R-phase-related peak was observed during the cooling run, commencing at approximately \SI{60}{\degreeCelsius}.

	To learn whether the variations of resistance were dominantly given by lattice properties or by changes in density of the charge carriers, field-induced Hall voltage was measured at selected temperatures during the cooling run in magnetic fields from \SI{-5}{\tesla} to \SI{5}{\tesla} and at DC currents of \SI{1}{\milli\ampere}, using the Physical Property Measurement System (PPMS-14, Quantum Design), leading to the determination of the Hall coefficient.
	The results are plotted together with the resistance curve in Fig.~\ref{fig:XrayMaps}b).
	The resistance peak at approximately \SI{40}{\degreeCelsius} does not correspond to any significant change of the Hall coefficient, that is, no change of the carrier density.
	Upon further cooling, however, the coefficient steeply increases indicating the onset of the transformation to martensite, consistent with the literature~\cite{Kunzmann2022}.

	\subsection{Elasticity and thermal diffusivity by TGS} 							\label{sec:TGS}

		\begin{figure}[t]
				\centering
				\includegraphics[width=\textwidth]{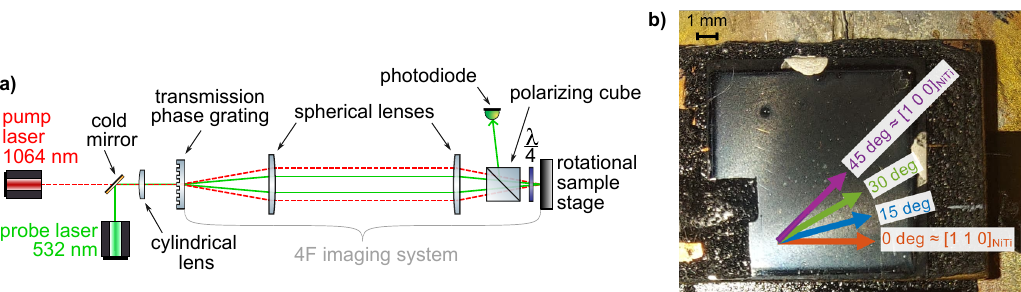}
				\caption{
					a) TGS optical setup (top view).
					b) The NiTi film mounted in the Peltier chamber, with the TGS measurement directions indicated.
					}
				\label{fig:TGSsetup_heatProp_sample}
		\end{figure}

	TGS was utilized to determine the elastic and thermal properties of the film.
	In the experimental arrangement---described in detail in Ref.~\cite{Stoklasova2021} and shown in Fig.~\ref{fig:TGSsetup_heatProp_sample}a)---the pump laser (\SI{0.5}{\nano\second} pulse duration, \SI{1064}{\nano\metre}) and the probe laser (continuous wave, \SI{532}{\nano\metre}) traveled along the same optical path, diffracting on a transmission phase mask and passing through a 4F imaging system before recombining at the sample surface to create spatially periodic interference patterns.
	The nominal fringe spacing was $\lambda = \SI{7}{\micro\metre}$, calibrated on a reference sample to $\lambda = \SI{7.07}{\micro\metre}$.
	In TGS, the excitation pattern generated by the pump laser acts as a spatially harmonic thermoacoustic source, inducing surface ripples combining transient thermal and acoustic gratings, as well as transient reflectivity changes.
	The time-domain dynamics of these gratings is then monitored by the diffracting probe beams, which are combined with the reflected ones in what is known as a differential heterodyne setup~\cite{Verstraeten2015,Stoklasova2021}, significantly improving the signal intensity and reducing noise.
	In this study, the signal was acquired by photodiodes, amplified in the bandwidth range from \SI{10}{\kilo\hertz} to \SI{1}{\giga\hertz}, and averaged over 50,000 waveforms in the time domain.

	The sample was housed within a custom-built chamber with a Peltier module for temperature control, and mounted on a rotation stage, which allows to measure the TGS signal at various in-plane directions on the sample surface.
	At \SI{120}{\degreeCelsius}, a high-resolution angular scan was performed, in which the sample was rotated in \SI{1}{\degree} steps, covering the range from $[1 1 0]_{\text{NiTi}}$ to $[1 0 0 ]_{\text{NiTi}}$.
	Then, a thermal cycle to \SI{5}{\degreeCelsius} and back to \SI{120}{\degreeCelsius} was performed, during which low-resolution angular scans were recorded every \SIrange{1}{10}{\degreeCelsius}, with a rotation step of \SI{15}{\degree}, as illustrated in Fig.~\ref{fig:TGSsetup_heatProp_sample}b).
	The thermal step varied based on proximity to the transformation temperatures, with finer steps near them.

	\medskip

		\begin{figure}[!t]
			\centering
			\includegraphics[width=\textwidth]{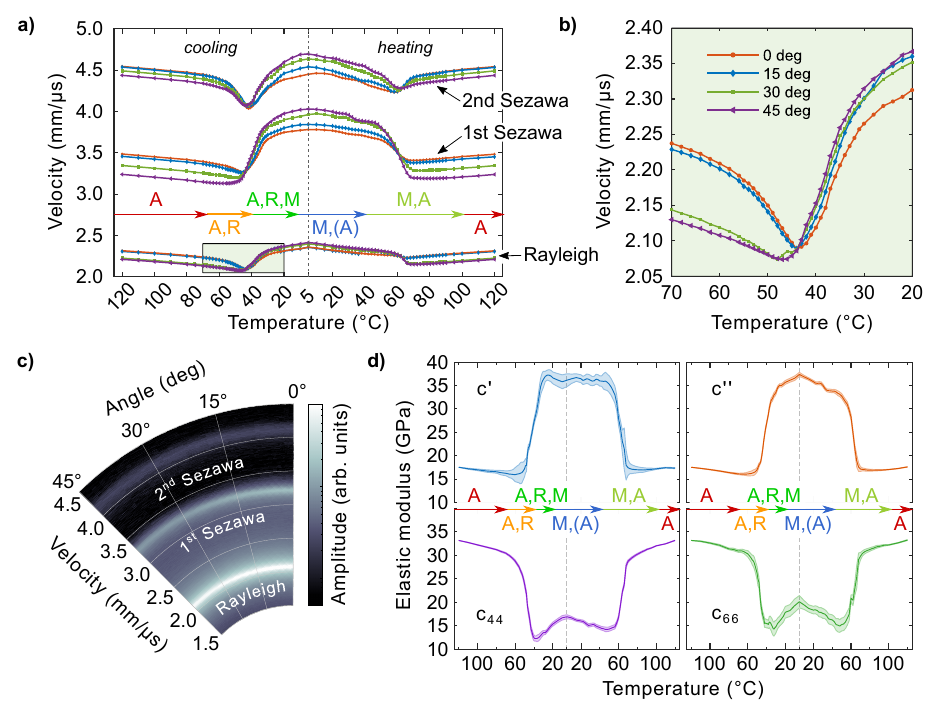}
			\caption{
				a) Evolution of Rayleigh and two Sezawa acoustic modes detected by TGS measurements during a temperature cycle of \SI{120}{\degreeCelsius} $\rightarrow$ \SI{5}{\degreeCelsius} $\rightarrow$ \SI{120}{\degreeCelsius}.
				b) Detail of the Rayleigh mode evolution upon cooling from \SI{70}{\degreeCelsius} to \SI{20}{\degreeCelsius} (as indicated by the green-background rectangle in the previous subfigure).
				c) TGS angular scan obtained at \SI{45}{\degreeCelsius}. Direction-independent velocity values for all detected acoustic modes indicate nearly isotropic behavior.
				d) Elastic coefficients (with corresponding error intervals) determined inversely from the velocity data.
				The colored arrows in a) and d) indicate the phase sequence as determined by XRD.
				}
			\label{fig:ResAcoustModesElConst}
		\end{figure}

	The acoustic response was analyzed by transforming the signal to frequency domain---more precisely, to velocity domain as $v=\lambda f$---with the resulting peaks corresponding to acoustic modes with prominent out-of-plane displacement, propagating along the surface; these are mainly Rayleigh and Sezawa modes.
	The velocity evolution of these modes in the selected directions is shown in Fig.~\ref{fig:ResAcoustModesElConst}a), with a detail of the Rayleigh mode in the temperature range of \SI{70}{\degreeCelsius} to \SI{20}{\degreeCelsius} shown in Fig.~\ref{fig:ResAcoustModesElConst}b).

	The angular dispersion of these modes was then used in an inverse calculation to determine the elastic coefficients of the material, described in detail in Ref.~\cite{Grabec2024}.
	The original high-resolution angular scan was used to precisely determine the elastic coefficients in the first step.
	During the thermal cycle, the elastic coefficients from the previous step were used as an initial guess for the inverse procedure.

	With the employed acoustic wavelength of $\SI{7}{\micro\metre}$, the acoustic modes propagate not only in the \SI{3}{\micro\metre} NiTi film, but are also influenced by the substrate.
	Therefore, the inverse calculation is based on a multilayer model, which accounts for the elastic properties of the substrate and buffer layers, as well as the film thickness.
		The MgO substrate with the buffer layers was treated as a single effective half-space, with the elastic coefficients determined previously for the same substrate~\cite{Grabec2024}.
	Furthermore, the density of the materials involved is required for the calculation of elastic coefficients from the velocity data.
	The effective density of the substrate with buffer layers was measured by Archimedes method as \SI{3.58}{\gram\per\cubic\centi\metre};
	the density of NiTi was considered to be \SI{6.45}{\gram\per\cubic\centi\metre}.
		Treating the buffers instead as separate layers with literature elastic coefficients and densities~\cite{Marklund1971,Levy2001} changes the determined elastic coefficients of the film by less than \SI{1.5}{\percent}.
	Although the changes in the elastic coefficients of the substrate in this temperature range were small ($\leq$\,\SI{3}{\percent}), they were accounted for in the analysis.

		Although the NiTi austenite is cubic, and self-accommodated martensite was shown to remain effectively cubic in its elastic behavior in bulk NiTi~\cite{Brill1991,Bodnarova2025}, the epitaxial constraint may break this symmetry; a tetragonal symmetry was therefore assumed throughout the thermal cycle.

		Since the detected acoustic modes are predominantly of shear character, they determine accurately the basal shears $c_{44}$ and $c_{66}$ and the diagonal shears $c^\prime = (c_{11} - c_{12})/2$ and $c^{\prime\prime} = (c_{11} + c_{12} - 4c_{13} + 2c_{33})/6$ (with respect to the cubic unit cell), which for cubic symmetry reduce to $c_{66} = c_{44}$ and $c^{\prime\prime} = c^\prime$.

	The thermal evolution of these coefficients is shown in Fig.~\ref{fig:ResAcoustModesElConst}d), and three significant states are highlighted in Tab.~\ref{tab:ElastConst}.
    Note that the error intervals of the elastic coefficients were estimated by the inverse procedure~\cite{Grabec2024}, and reflect the sensitivity of the acoustic modes to each coefficient rather than the overall experimental error.
	At all temperatures the elastic behavior was close to---but not perfectly---cubic, with $c_{44}\approx{}c_{66}$ and $c^\prime\approx{}c^{\prime\prime}$.
	Similar to observations of bulk single crystals~\cite{Brill1991,Bodnarova2025}, the basal shears ($c_{44}$ and $c_{66}$) are stiffer in the austenite phase, but undergo a pronounced softening, reaching a minimum upon the onset of the martensite formation.
	In contrast, the diagonal shears ($c^\prime$ and $c^{\prime\prime}$) undergo a significant stiffening once the martensite starts to form.
	Thus, the stiff and compliant shear modes switch their character during the transformation.

	This implies that there is a point during the transformation where the elastic behavior is isotropic, with $c_{44} = c_{66} = c^\prime = c^{\prime\prime}$.
	Indeed, such a point was found for the single crystals in the literature~\cite{Brill1991,Bodnarova2025}.
	Here, however, the perfect isotropy is not fully attained at any particular temperature, as the epitaxial connection to the substrate breaks the symmetry and allows $c_{44} \neq c_{66}$ throughout the whole transformation.
	The material was closest to isotropic at \SI{45}{\degreeCelsius}, where the velocities of all three observed acoustic modes became nearly independent of the in-plane direction of propagation, as shown in Fig.~\ref{fig:ResAcoustModesElConst}c).

		\begin{table}[!b]
			\centering
			\caption{Shear elastic coefficients of the NiTi film at three significant temperatures: in pure austenite and martensite phases, respectively, and at the temperature where the acoustic response is effectively isotropic. The uncertainties are the inversion errors from the sensitivity analysis described in~\cite{Grabec2024}.}
			\label{tab:ElastConst}
			\begin{tabular}{lrrrc}
				\noalign{\smallskip}
				\hline\noalign{\smallskip}
				& $c^\prime$ (GPa)& $c^{\prime\prime}$  (GPa)  & $c_{44}$ (GPa) & $c_{66}$ (GPa) \\
				\noalign{\smallskip}\hline\noalign{\smallskip}
				austenite (\SI{120}{\degreeCelsius}) 	 & $17.5 \pm 0.1$ & $17.5 \pm 0.1$  & $33.2 \pm 0.2$ & $33.2 \pm 0.2$ \\
				martensite (\SI{5}{\degreeCelsius}) 	 & $36.1 \pm 1.4 $ & $37.4 \pm 0.4 $  & $16.9 \pm 0.4 $  & $ 20.1 \pm 1.3 $  \\
				near-isotropic mixture (\SI{45}{\degreeCelsius})  & $ 22.8 \pm 0.5 $ & $ 21.3 \pm 0.2 $ & $ 15.6 \pm 0.6 $ & $ 17.3 \pm 0.8 $  \\
				\noalign{\smallskip}\hline
			\end{tabular}
		\end{table}

	\medskip

		\begin{figure}[!t]
			\centering
			\includegraphics[width=\textwidth]{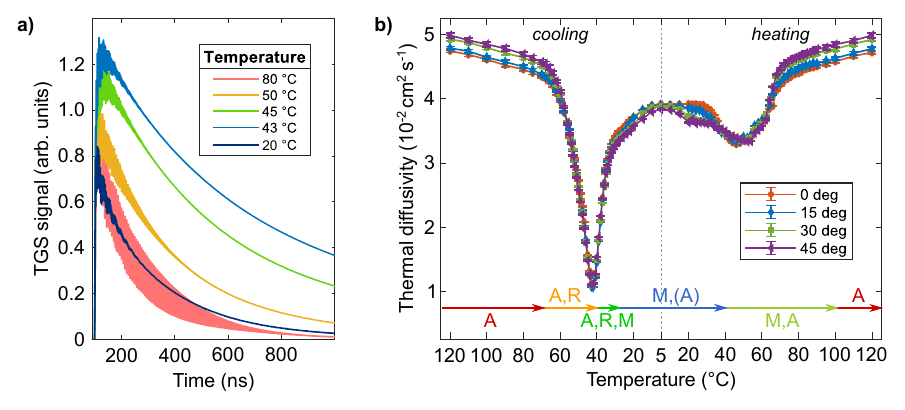}
			\caption{
				a) Selection of time-domain TGS signals recorded upon cooling from \SI{80}{\degreeCelsius} to \SI{20}{\degreeCelsius}, showing the heat-diffusion-related decay of the signal superimposed with the acoustic oscillations.
				b) Thermal diffusivity $\alpha$ determined from TGS signals through the full temperature cycle by fitting Eq.~\ref{eq:ThermDiff}; the slight directional spread is attributable to elastic anisotropy, discussed in~\cite{Kusnir2023}. Error bars (fit uncertainty) were below \SI{1}{\percent} and are smaller than the symbols.
				}
			\label{fig:ResTimeDomTGS_ThermDiff}
		\end{figure}

	Thermal diffusivity coefficient, $\alpha$, was determined by fitting the time-domain TGS signal with a model describing the decay of the surface ripple~\cite{Kading1995,Dennett2018}
		\begin{equation} \label{eq:ThermDiff}
			I(t) = A \left[\rm{erfc}\left(q\sqrt{\alpha t}\right) - \frac{\beta}{\sqrt{t}} \rm{exp}\left(-q^2 \alpha t\right) \right]
			+ B \sin \left(2\pi f t + \varphi\right) \exp\left(-\frac{t}
			{\tau}\right)
			+ C,
		\end{equation}
	where $q = 2\pi/\lambda$ is the grating wavevector, $\beta$ describes the ratio of displacement and reflectivity contributions, $\varphi$~is the phase of the acoustic signal, $\tau$~the acoustic decay constant, and $A$, $B$, and $C$~are amplitude constants.
		Frequency $f$ corresponds to the dominant peak in the TGS spectra, and its inclusion stabilizes the fitting~\cite{Dennett2018}.
	As the thermal properties are characterized through the homogenization of the surface temperature (and connected leveling of the surface ripple), the sensitivity of TGS to thermal properties is confined to a surface layer given by the heat diffusion length, approximately equal to $\lambda/\pi$~\cite{Kading1995}.
		Since in this case $\lambda/\pi \approx \SI{2.25}{\micro\metre}$, the chosen wavelength is sufficient for observing the heat flow through the \SI{3}{\micro\metre} thick NiTi film without significant contribution from the substrate and buffer layers.
		A finite-element simulation following~\cite{Kusnir2023}, in which the thermal conductivity of the substrate was varied by a factor of ten in both directions, changed the evaluated diffusivity by less than \SI{0.2}{\percent}.

	Figure \ref{fig:ResTimeDomTGS_ThermDiff}a) shows the time-domain TGS signals at selected temperatures during the cooling run.
	The decay of the signal is directly related to the thermal diffusivity, with a slower decay indicating a lower diffusivity.
	As shown in Fig.~\ref{fig:ResTimeDomTGS_ThermDiff}b), thermal diffusivity underwent a sharp decline beginning at the onset of R-phase transition (around \SI{70}{\degreeCelsius}).
	At \SI{43}{\degreeCelsius}---in the R-phase---the diffusivity reached its minimum of about \SI{1.1e-2}{\centi\metre\squared\per\second}, which was approximately \num{4.5} times lower than the value for austenite at \SI{120}{\degreeCelsius} (about \SI{4.9e-2}{\centi\metre\squared\per\second}, averaged over the measured directions).
	Once the martensite started to form, the diffusivity increased again, recovering to about \SI{80}{\percent} of the austenite value (compared with about \SI{22}{\percent} at the minimum).
	During the subsequent heating cycle---which proceeds directly from martensite to austenite---thermal diffusivity exhibited only a modest reduction.

	As also seen in Fig. \ref{fig:ResTimeDomTGS_ThermDiff}, the diffusivity values obtained by TGS experiments in different directions slightly differed, by approximately \SI{5}{\percent} at maximum.
	This difference is an artifact of the TGS data evaluation by Eq.~\ref{eq:ThermDiff}, which is derived for isotropic materials and does not account for the elastic anisotropy of the material, as discussed in detail in~\cite{Kusnir2023}.
	As demonstrated in~\cite{Kusnir2023}, the true value resides within the bounds defined by the extreme values.
	The uncertainty of $\alpha$ estimated from the fit of Eq.~\ref{eq:ThermDiff} was below \SI{1}{\percent} throughout the temperature cycle, both for the film and for the bulk crystal (Fig.~\ref{fig:bulk}); similarly to the elastic coefficients, it does not represent the overall experimental error.

\section{Discussion} \label{sec:Discussion}

	In the above summarized experimental data, the most pronounced feature is a thermal diffusivity drop by as much as a factor of \num{4.5} prior to the martensitic transition during the cooling run.

		Thermal diffusivity can be expressed as $\alpha = k/c_p \rho$, where $c_p$ is the specific heat capacity, $\rho$ the mass density, and $k$ the thermal conductivity.
		Since the mass density arguably stays constant during the transformation, we conclude that the enormous plummet in thermal diffusivity must be a consequence of changes in thermal conductivity or specific heat capacity.
		For metals, the thermal conductivity is typically proportional to the electrical conductivity via the Wiedemann--Franz law.
		As the electrical conductivity changes by less than \SI{10}{\percent} (see Fig.~\ref{fig:XrayMaps}b)), it is justified to consider similarly small changes also for $k$.
		Indeed, direct heat-pulse experiments~\cite{Ingale2011,Ramachandran2015,Ramachandran2017} reported similarly small changes in $k$ for Ni-rich and Fe- and Cu-alloyed NiTi materials at the R-phase transition temperatures, and no significant changes in $k$ were observed for free-standing NiTi thin films using $3\omega$ experiments~\cite{Jain2008}.
		This indicates that the enormous plummet in thermal diffusivity must be a consequence of changes in heat capacity.

	The rapid decrease in $\alpha$ upon cooling begins at around \SI{70}{\degreeCelsius}---where the R-phase starts to form according to XRD---
	{whereas the beginning of the martensitic transformation from the point of view of both the elastic coefficients and XRD data occurs between \SI{50}{\degreeCelsius} and \SI{40}{\degreeCelsius}, as shown in Figs.~ \ref{fig:ResAcoustModesElConst}d) and \ref{fig:XrayMaps}a), respectively.}
	This indicates that it is the R-phase formation, not the martensitic transition itself, that is responsible for the drop in thermal diffusivity.
	Consistently, the drop is absent during the heating run, where martensite transforms directly to austenite.
	The R-phase transition is known to be associated with a pronounced latent heat peak in differential scanning calorimetry (DSC, see e.g.~\cite{Duerig2015}).
	So is, however, the martensitic transition itself---but its effect on thermal diffusivity is much weaker.
	Thus, the heat capacity $c_p$, as extracted from the thermal diffusivity $\alpha$ determined from the TGS experiment, has a different meaning than the heat release/absorption rate of the progressing phase transition recorded by DSC.

	Importantly, the peak on the DSC curve related to R-phase during the cooling run is obtained under a constant exothermic heat flux.
	In the TGS experiment, by contrast, the pulsed laser acts as a periodic endothermic source superimposed with the cooling temperature ramp, making the TGS result more similar to temperature-modulated DSC (TM-DSC) than to classical DSC.

		However, using TM-DSC is not feasible for the studied sample:
		Considering the heat capacity of the substrate with respect to that of the film (the film accounts for only about \SI{2}{\percent} of the total), the contribution of the film to the DSC signal---even with its dramatic change---is below what can be reliably resolved, and removing the film from the substrate would alter its epitaxially constrained state.

	The laser energy, coming from the pump laser, does not drive the forward R-phase formation; rather, the nanosecond localized heating temporarily suppresses the rhombohedral distortion or atomic shuffle, which reappears once the thermal gradient dissipates and the temperature equilibrates.
	Such a mechanism is plausible only when the material undergoes a continuous, second-order-like transition, wherein an arbitrarily small temperature fluctuation induces a proportionally small change in the order parameter.
	Several studies suggest that the R-phase transition exhibits exactly this nature~\cite{Lukas2002,Miyazaki1988,Duerig2015}, in the sense that the rhombohedral angle (and, in turn, the lattice distortion) gradually evolves with temperature or applied loading.
	In contrast, if the transition proceeds via nucleation and growth of a daughter phase with significant thermal hysteresis, the minor thermal perturbations introduced by the pump laser are insufficient to drive cyclic phase transformations.
	As a result, the in-situ TGS measurement is insensitive to the ongoing macroscopic phase transformation, and the obtained value of $\alpha$ reflects solely the intrinsic properties of the instantaneous phase mixture, uncoupled from latent heat contributions.
	Thus, the continuous second-order R-phase transition and the strongly first-order martensitic transformation produce fundamentally different behavior of the TGS thermal diffusivity data.

		\begin{figure}[tbh!]
			\centering
			\includegraphics[width=0.5\textwidth]{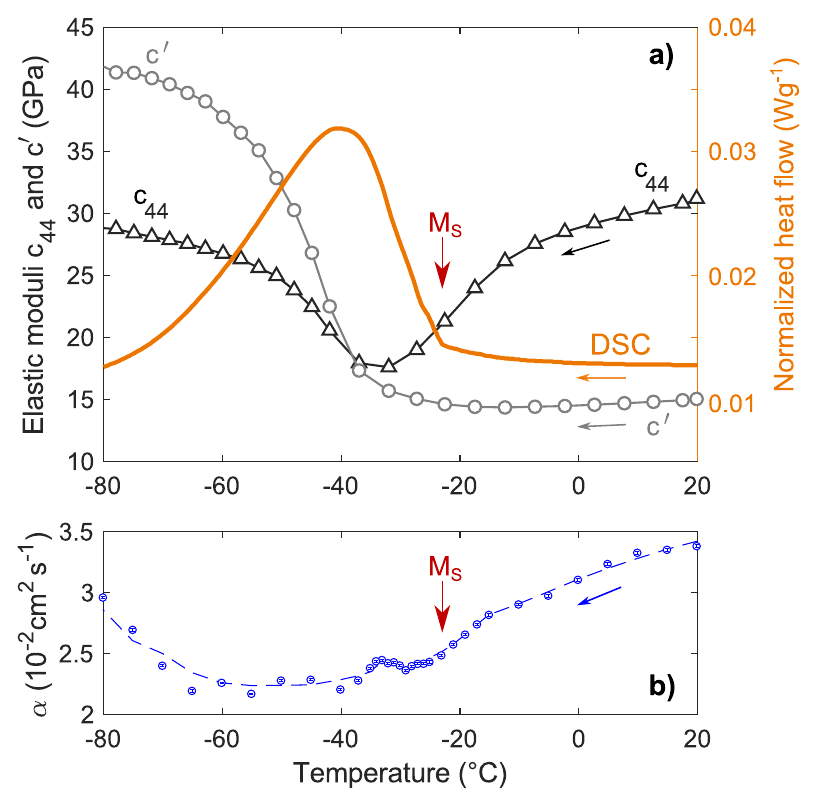}
			\caption{
				Complementary TGS experiment on a bulk single crystal~\cite{Bodnarova2025} exhibiting no detectable R-phase formation during the cooling run:
				a) a DSC curve (orange) showing only the martensitic peak and shear elasticity evolution (gray) determined by RUS,
				b) thermal diffusivity evolution by TGS (with small error bars showing fit uncertainty), showing no dramatic decrease near the transition.
				Notice also the generally lower thermal diffusivity
					 (about \SI{3.4e-2}{\centi\metre\squared\per\second} in austenite at \SI{20}{\degreeCelsius}, compared with about \SI{4.9e-2}{\centi\metre\squared\per\second} in the austenite of the film at \SI{120}{\degreeCelsius})
				in the single crystal with high volume fraction of TiC inclusions and Ti$_2$Ni precipitates compared with a perfect epitaxial film.}
			\label{fig:bulk}
		\end{figure}

	To establish whether this anomaly is specific to the R-phase, we performed an additional TGS experiment on the bulk single-crystalline sample of NiTi recently analyzed by resonant ultrasound spectroscopy (RUS) in Ref.~\cite{Bodnarova2025}.
	Unlike the film, this bulk sample does not undergo a two-step transformation through the R-phase, but a direct transformation from austenite to martensite (as confirmed by DSC; see Figure \ref{fig:bulk}a), where no R-phase related peak is seen).
	In line with the discussion above, the evaluated thermal diffusivity, shown in Fig.~\ref{fig:bulk}b), does not exhibit any pronounced drop.
	This directly confirms that the drop observed on the film is inherently connected to the presence of the R-phase.
	Notice, nevertheless, a shallow decrease in diffusivity associated with the martensitic transition in Fig.~\ref{fig:bulk}b)---similar in depth and shape to that observed in the epitaxial film during the heating run.
	This means that the variations of thermal diffusivity measured by TGS can be used to detect the martensitic transition temperatures, unless overlapped by the much stronger R-phase effect.

	Regarding the elasticity, the evolution for the bulk crystal (evaluated by RUS and re-drawn in Fig.~\ref{fig:bulk}a) is similar in character to that observed on the film: a pronounced drop of $c_{44}$ before the martensitic transformation, an isotropic point after its onset, and a switch of $c_{44}$ and $c^\prime$.
	Evidently, the evolution of elasticity alone cannot distinguish between the direct transformation and that going through R-phase.
	The softening of the basal shears likely acts as a precursor to the martensitic transition itself.
	The lattice modulations---specifically, the atomic shuffle and the rhombohedral distortion collectively identified as the R-phase transition---are simply alternative manifestations of this underlying mechanical instability in the parent austenite lattice.
	As recently discussed by Chen et al.~\cite{Chen2026}, the atomic shuffle and the rhombohedral distortion can occur independently or simultaneously; consequently, the $c_{44}$ softening may manifest with or without an accompanying macroscopic distortion.
	In the case of the investigated epitaxial film, the X-ray diffraction analysis confirms the presence of atomic shuffling, evidenced by the emergence of the $1/3\langle1\,1\,0\rangle^*$ superlattice reflections below \SI{70}{\degreeCelsius}.
	However, there is no distinct evidence of lattice parameter changes, implying that they are either negligibly small, or mechanically suppressed by the epitaxial constraints imposed by the substrate.

\section{Conclusions}											\label{sec:Conclusion}

We have demonstrated the ability of transient grating spectroscopy to simultaneously monitor thermal diffusivity and elastic coefficients of an epitaxial film.
When examining an epitaxial NiTi film as a model system, we observed a change of thermal diffusivity by a factor of \num{4.5} between the rhombohedral R-phase and cubic austenite.

This outstanding value makes NiTi a promising candidate for solid-state thermal switches, being of particular interest for thermal microsystems requiring compact, fast-switching solutions.

The practical viability of NiTi as a thermal-switch material will also depend on cycling stability and switching speed.
Epitaxial films, lacking large-angle grain boundaries, have shown substantially improved fatigue resistance in a related Heusler system~\cite{Fareed2024}.
Intrinsic transformation speed is unlikely to be limiting either: nanosecond-scale martensite-to-austenite switching has been demonstrated in a \SI{100}{\nano\metre} epitaxial NiTi film~\cite{Luenser2024}, though realizing comparable speeds at the \SI{3}{\micro\metre} thickness studied here will depend on thermal design rather than the material's intrinsic kinetics.

Furthermore, these huge changes in thermal diffusivity must be considered during design and simulations of any system exhibiting R-phase, as this intermediate phase can thermally shield the still untransformed austenite.

We attribute the observed drop in thermal diffusivity to the anomalous increase in heat capacity caused by the non-hysteretic gradual character of the R-phase transition.
In contrast, across the hysteretic, strongly first-order martensitic transition, the thermal diffusivity profile exhibits a significantly shallower and broader minimum.
The complementary TGS measurement on a bulk single-crystalline material without R-phase allows us to distinguish between two concurrent phenomena.
The instability of austenite preceding the martensitic transition manifests as elastic softening in $c_{44}$, while the R-phase transition manifests as an anomalous drop in thermal diffusivity.
The TGS experiment can simultaneously detect both these features and, consequently, help to distinguish between these two processes during the transformation.

\paragraph*{Acknowledgment}
This work has been financially supported by Czech Science Foundation [project No. 25-16285S] and Czech Ministry of Education, Youth and Sports [project No. CZ.02.01.01/00/22\_008/0004591]. Part of this work was supported by the German Research Foundation DFG under grant FA 453/13-1.

\section*{Author Declarations}

\paragraph*{Conflict of Interests} The authors have no conflicts to disclose.
\paragraph*{Author Contributions}
 K.~Rep\v{c}ek -- Investigation, Writing/Original Draft Preparation, Visualization; T.~Grabec -- Writing/Review \& Editing, Methodology, Conceptualization, Software; D.~Mare\v{s} -- Investigation, Data Curation, P.~Stoklasov\'{a} -- Methodology, Supervision; P.~Sedl\'{a}k -- Data Curation, Formal Analysis, Software; J.~Ku\v{s}n\'{i}r -- Investigation, Data Curation; P. Ve\v{r}t\'{a}t -- Investigation, Writing/Review \& Editing; O. Heczko -- Investigation; S.~F\"{a}hler -- Resources, Writing/Review \& Editing; K.~L\"{u}nser  -- Resources, Writing/Review \& Editing; H.~Seiner -- Conceptualization, Supervision, Project Administration, Writing/Review \& Editing, Visualization

\paragraph*{Data Availability Statement}
The data that support the findings of this study are openly available at the following URL/DOI: https://doi.org/10.5281/zenodo.20825245

\bibliographystyle{elsarticle-num}
\bibliography{references.bib}

\end{document}